\documentclass[aps,prl,preprint,superscriptaddress]{revtex4-1}
\usepackage{booktabs}
\usepackage{romannum}
\usepackage{graphicx}
\usepackage[export]{adjustbox}
\usepackage{subcaption}
\usepackage[hscale=0.85,vscale=0.85]{geometry}
\usepackage[font=footnotesize]{caption}
\usepackage{amsmath}
\usepackage{color}

\begin{document}


\title{Stochastic Resonances in a Distributed Genetic Broadcasting System: The $NF\kappa B/I\kappa B$ Paradigm}


\author{Zhipeng Wang}
\affiliation{Department of Statistics, Rice University, Houston, TX, 77005}
\affiliation{Department of Chemistry, Department of Physics and Astronomy, Center for Theoretical Biological Physics, Rice University, Houston, TX, 77005}
\author{Davit A Potoyan}
\author{Peter G Wolynes}
\affiliation{Department of Chemistry, Department of Physics and Astronomy, Center for Theoretical Biological Physics, Rice University, Houston, TX, 77005}



\begin{abstract}
Gene regulatory networks must relay information from extracellular signals to downstream genes in an efficient, timely and coherent manner. Many complex functional tasks such as the immune response require system-wide broadcasting of information not to one but to many genes carrying out distinct functions whose dynamical binding and unbinding characteristics are widely distributed.
In such broadcasting networks the intended target sites are also often dwarfed in number by the even more numerous non-functional binding sites.  Taking the genetic regulatory network of $NF\kappa B$ as an exemplary system we explore the impact of having numerous distributed sites on the stochastic dynamics of oscillatory broadcasting genetic networks pointing out  how resonances in binding cycles control the network's specificity and performance. We also show that active kinetic regulation of binding and unbinding through molecular stripping of DNA bound transcription factors can lead to a higher coherence of gene-co expression and synchronous clearance.  
\end{abstract}

\maketitle



\newpage 

 \section{Introduction}  
 

All living organisms must cope with dynamically changing environments. Cells of more complex organisms rely on sophisticated gene regulatory networks to rapidly and reliably acquire environmental information and communicate this information to downstream genes for action~\cite{davidson2001genomic}. The flow of information covers several time and length-scales starting from the diffusion limited encounter between segments of DNA and transcription factors and ranging up to cellular motion or cell death\cite{phillips2012physical}. Many master genes can broadcast information through cascades, oscillations, and waves of regulatory molecules to a wide range of genes within a cell and sometime even to neighboring cells~\cite{kruse2005oscillations}. In higher organisms these master genes broadcast signals to many downstream genes which must turn disparate biochemical processes on and off in synchrony with other genes~\cite{ZambranoElife}.

{ Here we explore how the resonances in temporal patterns of non-equilibrium binding and unbinding processes to disparate binding sites on the genome are amplified cooperatively for function. The difficulty of achieving synchrony in turning genes on or off has been ignored in most models of gene regulation, which generally assume ultra fast binding equilibration at genomic sites~\cite{bintu2005transcriptional}. These models based on binding equilibrium would imply that there must be a large binding free energy gap separating the affinities of transcription factors for target sites and for non-functional sites in order to ensure proper function of a broadcasting network. This thermodynamic distinction need not be true for genetic networks which are operating far from equilibrium. To highlight how the stochastic dynamics of binding and unbinding to numerous genomic sites has non-trivial dynamic consequences we study an example of  a broadcasting network that involves the transcription factor $NF\kappa B$~\cite{HoffmannScience,NelsonScience}. This transcription factor regulates hundreds of genes and binds strongly to a myriad of genomic sites. The $NF\kappa B$ broadcasting  network crucially contains a time-delayed negative feedback loop, where $NF\kappa B$ induces the transcription of its own inhibitor $I\kappa B$ (Fig 1A). This inhibitor ultimately restores a quiescent steady state by clearing $NF\kappa B$ from the nucleus by binding to $NF\kappa B$ and translocating $NF\kappa B$ into the cytoplasm where the $NF\kappa B$ now bound to the inhibitor will wait until new stimuli from the environment are encountered. Under continuous stimulation the $NF\kappa B$/$I\kappa B$ network exhibits sustained oscillatory dynamics~\cite{HoffmannScience,NelsonScience,SneppenPNAS}. In addition to the $I\kappa B$ response, which leads to the pulses and oscillations,  numerous binding sites capture and release the $NF\kappa B$ with widely varying rates. The non-linear influence of the target and decoy sites on the rest of the networks is indirect, arising ultimately because all the sites need to access a shared resource, the $NF\kappa B$ signal molecule itself. 

In thermodynamic models~\cite{StormoZhao,bintu2005transcriptional} the local DNA sequence determines the probability of a transcription factor being bound to the site through the free energy of binding $\Delta F_b = {k_B T} \ln (k_{f}/k_{b})$, where $k_{b}$ and $k_{f}$  are the binding and unbinding rate coefficients respectively. While it has also been assumed that in eukaryotic gene regulation the target binding sites possess much higher binding affinities than any random sequences, recent protein-binding microarray(PBM) experiments of NF$\kappa$B proteins\cite{TSiggersNatureImmunology} show that the binding affinities of NF$\kappa$B to target binding sites are distributed over a wide range (Fig. 1C).  Some functional binding sites for NF$\kappa$B even have binding affinities that are weaker than the affinity of random sequences for $NF\kappa B$. The recent discovery of induced molecular stripping whereby the inhibitor $I\kappa B$ can irreversibly strip $NF\kappa B$  from genomic sites to which it binds further defies the generality of purely thermodynamic thinking \cite{AlverdiPNAS,potoyan2016molecular}.  The non-equilibrium nature of gene regulation in an oscillating broadcasting network thus demands a comprehensive kinetic model which takes into account the heterogeneous distribution of binding/unbinding rates for genomic sites and the consequences of the distribution of rates for the expression of target genes. 

{From the phenomenological point of view the heterogeneity of unbinding rates at different $DNA$ sites creates a distribution of stochastic limit cycles with varying oscillation periods. These cycles of binding and unbinding to genome wide DNA sites would naturally oscillate out of phase with respect to each other.  Since each binding site which is associated with a particular gene is a single molecule entity, the stochastic nature of expression for the downstream signaling pathways would be further amplified~\cite{PotoyanPNAS}. The fact that the large number of incoming $I\kappa B$ molecules  can directly remove $NF\kappa B$ from its binding sites creates a possibility of enhancing the coherence of these oscillators via a mechanism of cooperative dissociation. To what extent this coherence enhancement is exploited in real cells needs to be tested in experiments by engineering reporter gene assays with different binding affinities and monitoring their oscillatory patterns with single cell resolution. In the synthetic biology context designing ways to enhance coherence or reducing the dichotomous noise of promoter binding is one of the main challenges for modular design of functional circuits~\cite{jayanthi2012tuning}. So far this problem of engineering modular circuits has been combatted in a brute force fashion by increasing  the cooperative association of proteins with $DNA$ or by adding more copies of the promoter sites~\cite{jayanthi2012tuning,KarapetyanPRE,lengyel2017multiple}.}
    
\setcounter{figure}{0}
\begin{figure}[!ht]
\centering
\includegraphics[width=0.8\textwidth]{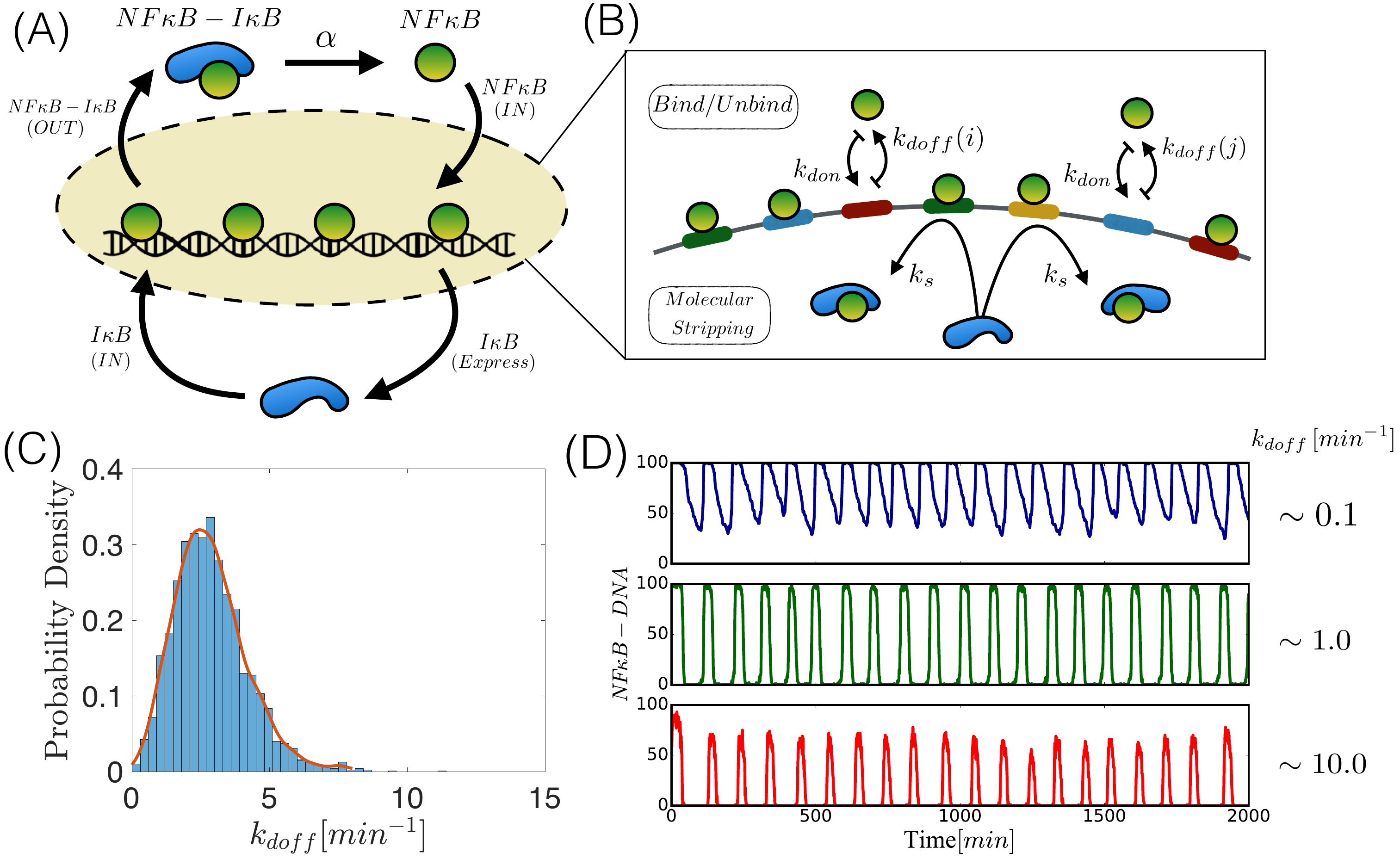}
 \caption{(A) Schematic view of the $NF\kappa B$ regulatory cycle showing the key steps of $I\kappa B$ production translocation in and out of the nucleus involving sequestration of $N\kappa B$ back to cytoplasm. (B) Two distinct mechanisms of interaction of $NF\kappa B$ with the genomic sites: passive binding/unbinding and molecular stripping (C) Distribution of unbinding rates, $k_{doff}$ of $NF\kappa B$ from DNA binding sites inferred from Protein Binding Microarrays (PBMs) experiments \cite{TSiggersNatureImmunology}. (D) Shown are stochastic trajectories of 100 genomic sites of $NF\kappa B$ sampled from the histogram distribution in 1C.}
 \label{fig:3}
\end{figure}

\section{Modeling broadcasting network of $NF\kappa B$ with distributed genomic binding sites.}
The stochastic dynamics of the $NF\kappa B$ broadcasting network in the well stirred limit is governed by a master equation which relates the change of probability for a particular micro-state of the network to changes in the numbers of molecules, $z$ as well as the occupancy state of the genomic binding sites, $\sigma$,  where $z = \{z_1, z_2, ..., z_N \}$ is a vector containing the numbers of molecules of each of the $N$ chemical species in the network, and $\sigma \in \{0, 1 \}$ is the binary variable representing the occupancy state of the genomic binding sites, with 0 indicating unoccupied state and 1 indicating occupied state. 
\begin{equation}
\begin{split} 
 \dot P(z,\sigma) &= J_{birth/death}(z \pm 1 \rightarrow z,\sigma) - J_{birth/death}(z \rightarrow z \pm 1,\sigma) + \\ 
 & J_{bind/unbind}(z \rightarrow z', \sigma \rightarrow \sigma')-  J_{bind/unbind}(z'\rightarrow z, \sigma' \rightarrow \sigma)
\end{split}              
\end{equation} 

In this equation the first two terms ($J_{birth/death}$) denote the ingoing and outgoing probability fluxes via birth/death processes that change the total number of molecules (z) while the last two terms ($ J_{bind/unbind}$) stand for probability fluxes caused by changes in the binary state ($\sigma$) of the binding sites (ON/OFF or bound/unbound). We employ a kinetic Monte Carlo scheme for solving  the master equation of broadcasting network\cite{GillespieJPC} accounting for all of the discrete changes in the numbers of states of genomic binding sites (Supplementary Tables 1 and 2). The total number of $NF\kappa B$ molecules in a cell is $10^5$ and is kept constant by setting its degradation rate to zero~\cite{biggin2011animal}. This is a reasonable approximation because $NF\kappa B$ is known to have a very long cellular lifetime~\cite{biggin2011animal}. We can estimate the number of genomic binding sites using genome-wide Chip-seq assays of binding, which have detected more than $2\times 10^4$ distinct DNA sites that bind to $NF\kappa B$ with an equilibrium affinity comparable to that for the $I\kappa B$ promoter \cite{BenZhaoChipSeq}. We call all of these $NF\kappa B$ genomic binding sites except for the $I\kappa B$ promoter "decoys" since the way that both specific and non-specific sites affect the stochastic dynamics of main regulatory loop of $NF-\kappa B/I\kappa B$ is by sequestering the free molecules of $NF\kappa B$. In other words we assume that there is no further feedback from downstream signaling processes triggered by $NF\kappa B$ binding to specific sites. The cell volume is set to 100 $\mu m^3$ consistent with the range of eukaryotic cell size. We assume a normal distribution for the binding free energies: $\Delta G_b \sim \mathcal{N}(\Delta \bar{G}, \bar{\sigma}^2)$. Transcription factor binding to DNA is commonly thought to be diffusion limited so we can assume fast and uniform binding ON rates of $NF\kappa B$ both to the $I\kappa B$ promoter and to all of the other binding sites ($k_{on} = k_{don} = 10 \mu M^{-1} min^{-1}$). The heterogeneity in binding free energies as measured by Chip-seq and microarray experiments leads then to strong heterogeneity in the unbinding rates $k_{doff}$ (Fig 1C). For the $I\kappa B$ promoter site, which sets up the oscillations the unbinding OFF rate $k_{off}$ is set to $1 min^{-1}$. This value generates an oscillation period consistent with single cell experiments in HeLa cells\cite{NelsonScience}. The unbinding rates $k_{doff}$ for different genomic sites takes on a log-normal distribution under the assumption that binding free energy follows a normal distribution: $\ln k_{doff} \sim \mathcal{N} (\Delta \hat{G},\sigma^2)$, where $\Delta \hat{G} = \frac{\Delta \bar{G}}{k_{B}T} + \ln k_{don}$ and $\sigma^2 = (1/k_{B}T)^2 \bar{\sigma}^2$.  In order to perform Monte Carlo simulations for the stochastic model, we approximate the log-normal distribution of $k_{doff}$ using the histogram probability density estimator \cite{ScottHistogram} (See Supplementary Materials Part I), and then, crucially use composite population variables. Thus we approximate continuous distribution of unbinding rates for thousands of genomic sites by grouping sites with closely spaced $k_{doff}$ values into $\sim 15-20$ histogram bins. Each bin can then bind a certain finite  number of transcription factors. Without making this grouping, direct stochastic simulation (at the individual site level) would be infeasible for realistically large eukaryotic genomes. By varying both $\Delta \hat{G}$ and $\sigma^2$ we investigate how the heterogeneity in the unbinding rate distribution affects the dynamical characteristics of the network.  \\

\section{Results and Discussions}

\indent To quantify the temporal coherence of the oscillatory dynamics we calculate the normalized autocorrelation function of both the number of bound $NF\kappa B$ molecules and the number of free molecules of $NF\kappa B$. We quantify the loss of  temporal coherence by calculating the dephasing time ($\tau_{\phi}$) of an exponential decay ($e^{-t/\tau_{\phi}}$) fitted to the envelope of a periodic $[cos(2\pi t/T)]$ normalized autocorrelation function (Supplementary Figure S3). Following the protocol that was used  in previous works \cite{PotoyanPNAS,KarapetyanPRE,MorelliPRL}, we use the oscillation quality to characterize the temporal coherence which is the ratio of the dephasing time to the oscillation period: $\tau_{\phi}/T$. A larger value of the oscillation quality indicates higher temporal coherence. Fig 2A illustrates the oscillation quality of the total amount of $NF\kappa B-DNA$ complex  as a function of $\Delta \hat{G}$ and $\sigma^2$, in the absence of molecular stripping ($k_{s} = 0 \mu M^{-1} min^{-1}$) and in the presence of active molecular stripping ($k_{s} = 10 \mu M^{-1} min^{-1}$). 

First, by considering the case of uniform binding sites ($\sigma^2=0$) we find that when the time-scale of $NF\kappa B$ unbinding from genomic sites $\tau_{doff}=k^{-1}_{doff}$ is comparable to the time-scale of unbinding to the $I\kappa B$ promoter $\tau_{off}=k^{-1}_{off} \sim 1 min$ these sites enter a phase of "resonance"  which manifests itself in the modulation of the oscillatory characteristics of the main cycle (3A-C). In particular these resonant sites contribute to the shortening of the oscillation period (Fig 3B) as well as the global coherence in oscillations of both DNA bound $NF-\kappa B$ molecules (Fig 3A) and oscillations of all the species in the the main loop (mRNA, IkB, $NF\kappa B$, Fig 1, SI).  In contrast, for values of unbinding rates that fall far outside of the "resonance" range, the main $I\kappa B$ oscillator is unaffected in its coherence and retains its uncoupled period $\tau_{osc}\sim 2hr$ (Fig 3B). The different manner in which fully resonant and fully non-resonant binding sites contribute to the oscillatory network leads to non-trivial behavior in the scenario where there are distributed $DNA$ binding sites. For finite values of $\sigma^2$ the shape of the distribution of the unbinding rates leads to distinct patterns of temporal coherence (Fig 2A). When the mean of the distribution of unbinding rates is comparable to the unbinding rate of promoter $\langle k_{doff} \rangle \sim k_{off}$, increasing the heterogeneity of rates $\sigma^2$ results in greater temporal coherence (Fig 2A). At first, this result, indicating that heterogeneous sites  oscillate more in phase than the do those sites having exactly the same rate may appear counterintuitive. This greater coherence, however is explained simply by noting that increasing $\sigma^2$ reduces the weight of the "resonant" sites in the distribution, which results in overall elevated temporal coherence of the oscillations. When the mean unbinding rate $\langle {k}_{doff} \rangle$ of the sites lies far outside of the resonance region, temporal coherence becomes relatively insensitive to the shape of the distribution. Enabling a process like molecular stripping, enforces cooperative dissociation of the $NF\kappa B$ from numerous sites via rapid $I\kappa B$ titration which is independent of the value of unbinding rates. This cooperative dissociation results  significant enhancement of the coherence of oscillations (Fig 2A, 3D). 

Here we use the term stochastic resonance in its most liberal sense to emphasize that molecular stochasticity associated with genome wide distribution of unbinding rates can modulate oscillatory dynamics of genetic broadcasting networks. Precisely how broadcasting gene networks leverage this resonance like property for functional purposes is unclear. Oscillatory networks may benefit from having sites with widely distributed unbinding rates (as opposed to uniform rates) so as to minimize resonance effects from genomic sites which would interfere with the main negative loop and reduce the coherence of $NF\kappa B/I\kappa B$ oscillations. It is also possible that broadcasting networks specifically target those resonant sites which carry out important functions and whose downstream activity is detected and regulated by the degree of coherence of oscillations. In a more general sense our simulations suggest that for networks with transcription factors that have access to a large number of genomic sites, the distributed nature of unbinding rates (or residence times) plays an important role for the temporal dynamics of the regulatory network as a whole. The specific functional roles for the temporal resonances associated with distributed genomic binding remain to be clarified in future experiments. We note that similar resonance like phenomena have been observed in other biochemical networks.   Lan and Papoian have found an instance of stochastic resonance in the dynamics of enzymatic signaling cascades~\cite{lan2007stochastic} where certain optimal rates in the sequence of enzymatic reactions lead to faster propagation of signals. Bates et all~\cite{bates2014stochastic} have found that optimal levels of noise enhance the learning performance of intracellular genetic perceptrons~\cite{bates2014stochastic}. One can not rule out that such resonance phenomena may sometimes also have deterministic origins. For instance, the bifurcation analysis of a purely deterministic model of a simple self-repressing genetic oscillator in the fast gene switching regime uncovered a different resonance like phenomenon~\cite{morant2009oscillations}: certain intermediate values for promoter state relaxation were optimal for destabilizing the steady states and enhancing the oscillations. The resonance phenomenon in the broadcasting network considered here provides both a non-linear and a stochastic component because both the shape of the limit cycles (Fig 3CD) as well as their transverse fluctuations (Supplementary Fig S7-8) are affected by the presence of resonant sites which taken together lead to shifting and broadening the period of oscillations in the main cycle. \\

%


\setcounter{figure}{1}
\begin{figure}[!ht]
\centering
\includegraphics[width=0.7\textwidth]{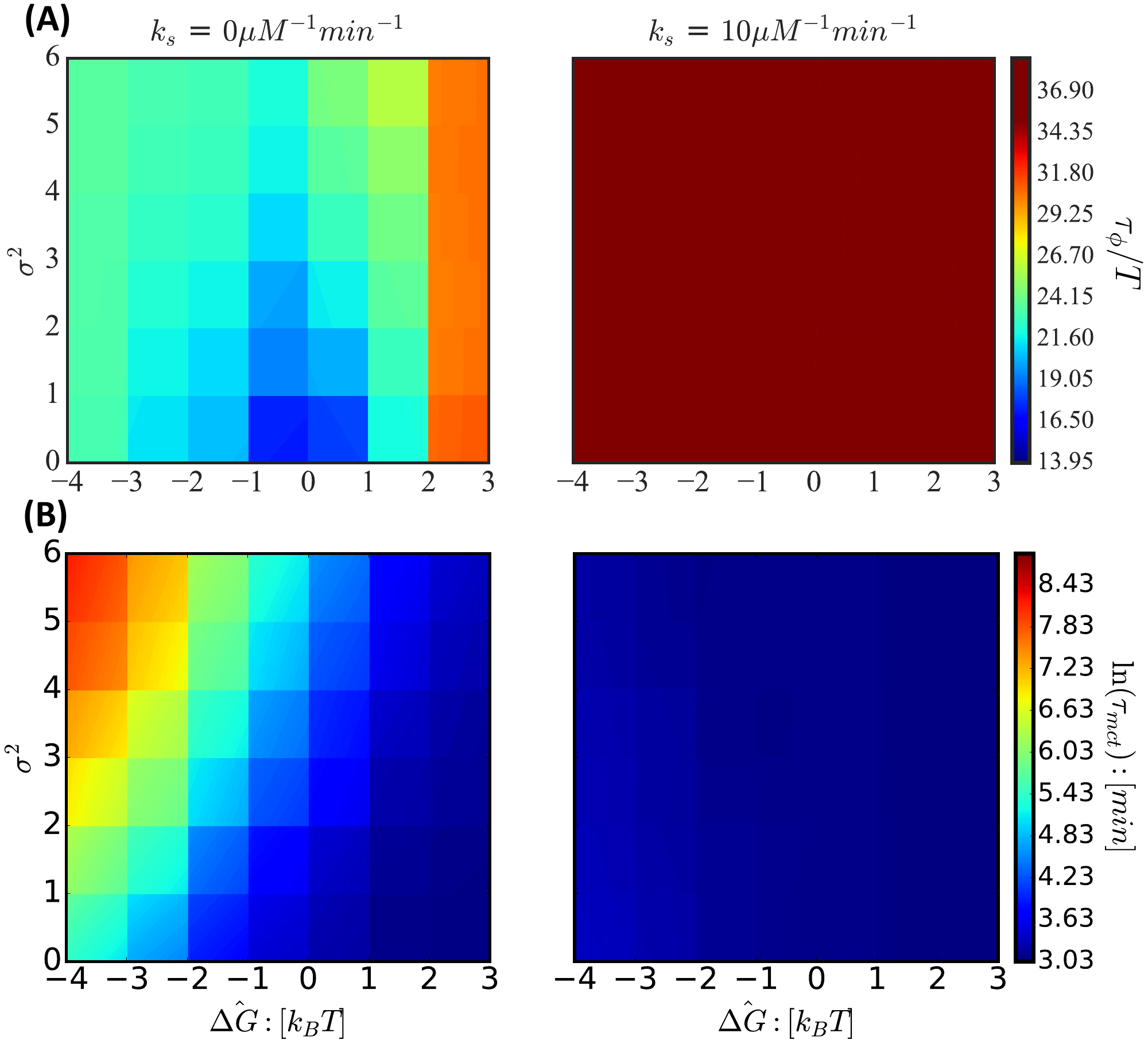}
 \caption{(A) Oscillation Quality $(\tau_{\phi}/T)$  as a function of mean ($\Delta \hat{G}$) and standard deviation ($\sigma$) of the distribution of unbinding rates from $2\times 10^4$ $NF\kappa B$ decoys (genomic sites except for the $I\kappa B$ promoter) when molecular stripping is disabled ($k_s = 0\, \mu M^{-1} min^{-1}$) and with molecular stripping enabled ($k_s = 10\, \mu M^{-1} min^{-1}$) (B) Log of mean clearance time $\ln (\tau_{mct})$ as a function of mean ($\Delta \hat{G}$) and standard deviation of distribution of unbinding rates from $2\times 10^4$ $NF\kappa B$ decoys when molecular stripping is disabled ($k_s = 0\, \mu M^{-1} min^{-1}$) and with molecular stripping enabled ($k_s = 10\, \mu M^{-1} min^{-1}$). }\label{fig:3}
\end{figure}

\indent When stimulation of the network is abruptly terminated by setting the degradation rate to zero $\alpha=0$ the binding sites will eventually be cleared of $NF\kappa B$. The relevant time for a meaningful response then is the time for all  the $NF\kappa B$ molecules to be removed from the bound sites. This time would determine whether some downstream genes would remain on when they should have been turned off, as the stimulus has aborted.  We calculate the mean clearance time (MCT) of bound decoys $\tau_{mct}$, which is the time it takes on average for all of the bound $NF\kappa B$ molecules  to dissociate from genomic sites. After termination of the signal, the number of $NF\kappa B$ molecules undergoes a random walk where each molecule independently dissociates from its $DNA$ binding sites but can later re-bind to the unoccupied sites during the attempt of the system to be cleared of genetically bound $NF\kappa B$. The master equation for the time-evolution of probability of the number of DNA bound $NF\kappa B$ molecules is: 

\begin{equation}
  \dot P(n,t) = k_{don} n_F P(n | n-1) - k_{doff}(n) P(n-1 |n,t) - k_{s}n_I P(n-1 | n )
\end{equation}

{where $n$ is the number of $NF\kappa B$ molecules bound to the genomic sites, $n_F$ is the total number of free nuclear $NF\kappa B$ molecules ($n_F+n\approx 10^5$),  $k_{don}, k_{doff}$(n) are the binding/unbinding rates from the genomic sites, $k_s$ is the molecular stripping rate ($I\kappa B+NF\kappa B\textendash DNA \xrightarrow{k_s} DNA + NF\kappa B\textendash I\kappa B$) and $n_I$ is the number of nuclear $I\kappa B$ molecules. This is a coarse grained equation obtained from Eq 2, by retaining the last two terms of Eq 1 accounting for the binding/unbinding and stripping only, which is justified due to rapid and irreversible nature of $NF\kappa B\textendash I\kappa B$ translocation into cytoplasm and slow rate of  $NF\kappa B\textendash I\kappa B$ dissociation while in nucleus (Supplementary Materials).} The Eq 2 is solved with reflecting and absorbing boundary conditions placed at $n=N$ and $n=0$ respectively with $N\sim 10^4$. The mean clearance time $\tau_{mct}$ can then be calculated as: $\tau_{mct} = \int_{0}^{\infty} t P(0,t | N,0) dt $. When molecular stripping is disabled the problem is similar to a 1D random walk on a lattice~\cite{bar1998mean} allowing us to estimate $\tau_{mct}(N)$, scales like $\sim e^N$ when rebinding is dominant, i.e. when $k_{don}n_F \gg k_{doff}$ and in contrast $\tau_{mct}(N) = \sum^{N-1}_{n=0} \frac{1}{k_{doff}(n)(N-n)}$ when rebinding is negligible i.e. $k_{don}n_F \ll k_{doff}$. When Molecular stripping is enabled the last term in Eq 2 dominates due to a large influx of $I\kappa B$ which leads to equally high $k_{doff}(n) \sim k_s n_I$ rates, thereby making the translocation of the $I\kappa B$ into the nucleus the rate-limiting step for the clearance. As the numerical solution of the master equation shows (Fig. 2B) the mean clearance time indeed depends strongly on the distribution of unbinding rates when molecular stripping is disabled. Sites with slower unbinding rates from $DNA$ sites naturally contribute the most to having a longer mean clearance time. The order of magnitude of the clearance time is largely dominated by the small number of sites that have the slowest dissociation rates in the distribution. When the molecular stripping mechanism is enabled, the cooperative dissociation of $NF\kappa B$ leads to much more rapid and uniform distribution of clearance times which now become insensitive to the heterogeneity characterized by the unbinding rate distribution. Molecular stripping is thus able to actively enforce coherence for all of the sites (Fig. 3).\\
 
\begin{figure}[h!]
\centering
\includegraphics[width=0.7\linewidth]{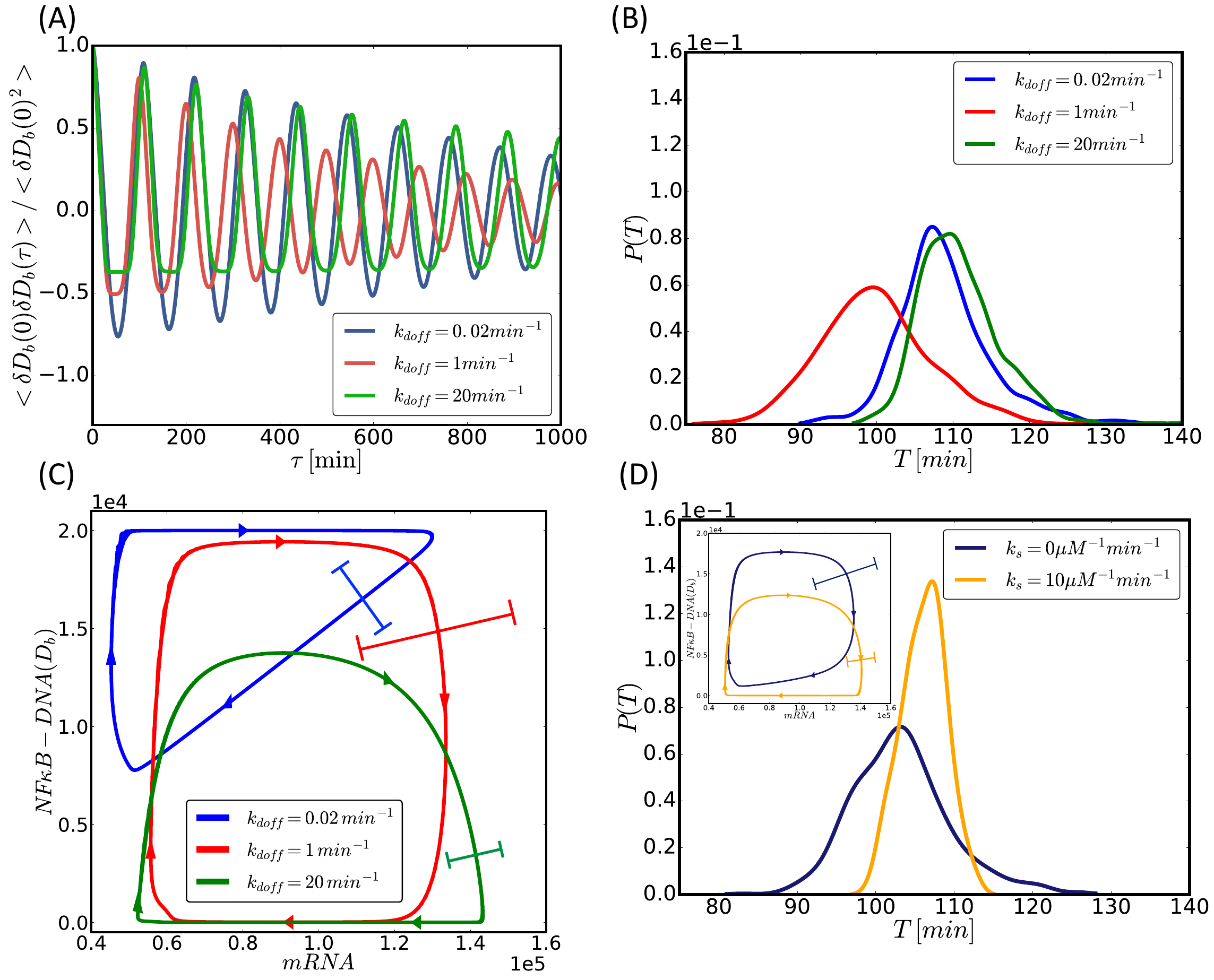}
  \caption{(A) Normalized autocorrelation of $NF\kappa B-DNA (D_b)$ population for $2\times 10^4$ identical $NF\kappa B$ decoys ($NF\kappa B$ genomic sites except for the $I\kappa B$ promoter) with different unbinding rates (B) Distribution of oscillatory period ($T$) of the $NF\kappa B-DNA (D_b)$ population for different unbinding rates of identical $NF\kappa B$ decoys (Number = $2\times 10^4$)  (C) \textbf{Limit cycles in the phase space of ensemble averaged numbers of $NF\kappa B-DNA (D_b)$ and mRNA species, with different unbinding rates of identical $NF\kappa B$ decoys (Number = $2\times 10^4$). The error bars show the noise level associated with each limit cycle}.  (D) \textbf{Distribution of oscillatory period ($T$) of $NF\kappa B-DNA (D_b)$ population for the case of $2\times 10^4$ distributed $NF\kappa B$ decoys whose unbinding rates follow a log-normal distribution with mean $\ln k_{doff} =0$ and $\sigma^2 = 6$. Two scenarios are presented one in the  molecular stripping (dark blue lines) and with molecular stripping (orange lines). The inset shows the limit cycles as a function of numbers of  $NF\kappa B-DNA (D_b)$ and mRNA species}.}\label{fig:3}
\end{figure}




\begin{figure}[h!]
\centering
\includegraphics[width=0.7\linewidth]{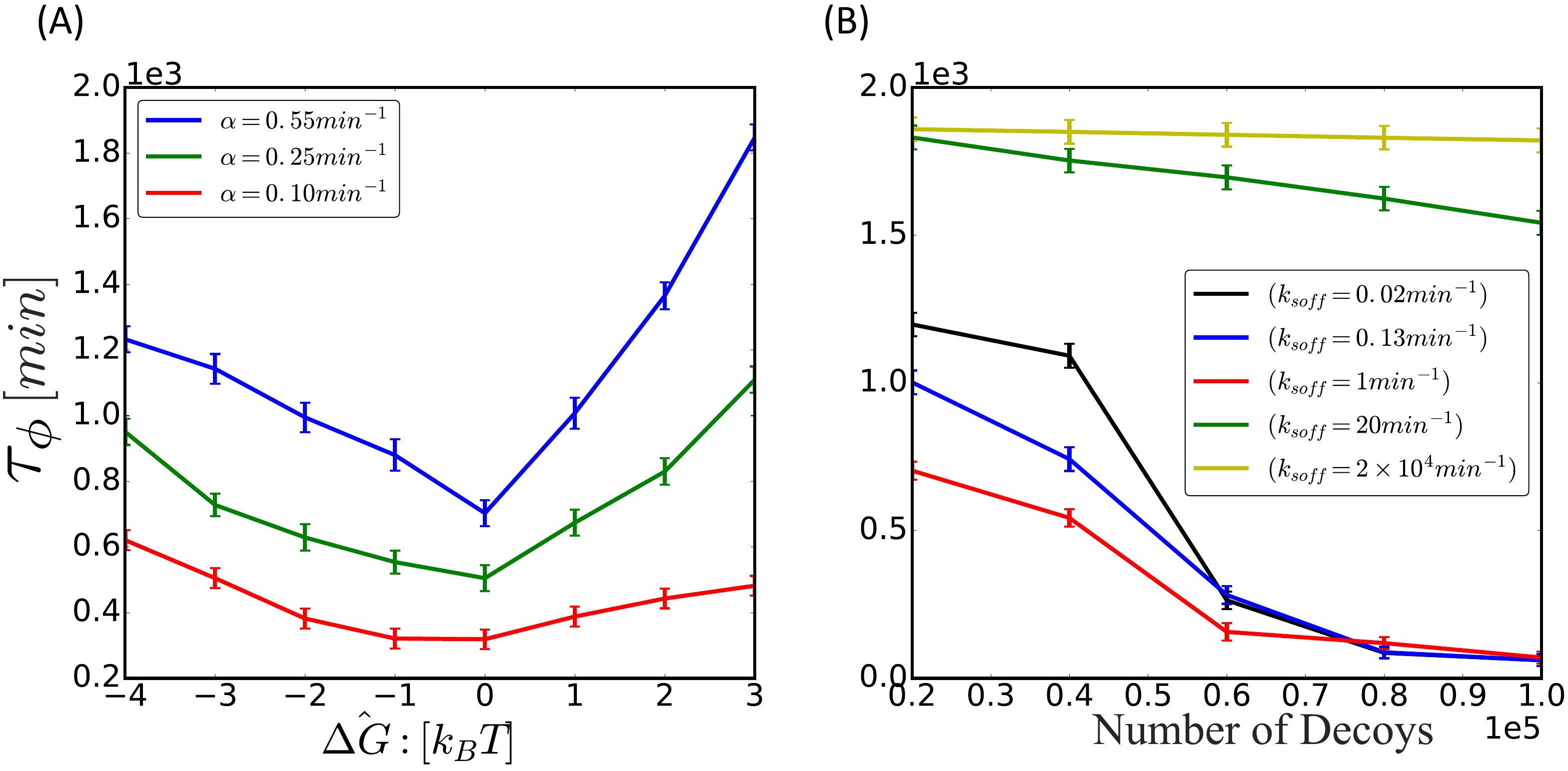}
  \caption{(A) Dephasing time $(\tau_{\phi})$ in the absence of molecular stripping as a function of $\Delta \hat{G}$  for three different levels of stimulation intensity ($\alpha = 0.55, 0.25, 0.10\,min^{-1}$). $2\times 10^4$ identical $NF\kappa B$ decoys ($NF\kappa B$ genomic sites except for the $I\kappa B$ promoter) are included in the system ($\sigma^2 = 0$) (B) Dephasing time $(\tau_{\phi})$ in the absence of molecular stripping as a function of the total number of identical $NF\kappa B$ decoys with different unbinding rates.}\label{fig:3}
\end{figure}


\indent We have also investigated the dependence of the resonance effects on stimulation intensity $\alpha$(Fig 4A) which quantifies the degradation rate of $I\kappa B$ in the $NF\kappa B-I\kappa B$ complex encoded by the input signals. We find that resonance like effects diminish as $\alpha$ tends to zero and system approaches the non-oscillatory clearance regime. Many much weaker binding sites that go unnoticed by  Chip-seq certainly exist on the genome. These could, in principle, change the resonance picture. To address this  possibility we carried out additional simulations by adding an extremely  large number of genomic sites with very disparate unbinding rates (Fig 4B). We pick five different decoy unbinding rate distributions for these extra sites, covering a very slow unbinding rate ($k_{doff} = 0.02 min^{-1}$), a resonant regime and a rather fast unbinding rate ($k_{doff} = 2 \times 10^4 min^{-1}$) for the thermodynamically weak additional decoys.  The temporal coherence of the network is most sensitive to the addition of decoys with unbinding rates that fall in resonant regime but the coherence is relatively  insensitive to adding either slower or faster unbinding sites. Once the decoy unbinding rate becomes very fast ($k_{doff} = 2\times 10^4 min^{-1}$), increasing the number of decoys has a nearly negligible effect on the temporal coherence. This insensitivity could potentially explain why the non-specific binding of transcription factors like $NF\kappa B$ to a large portion of the enormous eukaryotic genome does not affect the dynamics of even a very extensive genetic broadcasting network. \\

{ Our work shows that genome wide distributed affinities of binding sites can encode distinct dynamic regulatory information which can be parsed more clearly via oscillatory signals. Taking the $NF\kappa B$ circuit as an example we demonstrate that a number of special sites (which we call resonant sites) whose time-scale matches that of the main broadcasting site (The $I \kappa B$ promoter site) can become strongly coupled to the main cycle while others remain uncoupled and dynamically irrelevant. These resonant sites shift the oscillatory period of the main cycle thereby providing a specific feedback mechanism. For instance  the greater the number of sites that are available to tune in to the broadcasting of the $I\kappa B$ oscillatory signal the more its period is shifted. We have also shown that specifically the $NF\kappa B$ network in mammalian cells can cope with having tens of thousands of non-specific binding sites that are known to be present by using molecular stripping like mechanisms. Such a mechanism enables a rapid turnover of $NF\kappa B$, despite the existence of numerous binding traps, which, in turn, leads to coherence in gene co-expression when the stimulus is on and rapid co-clearance of sites when the stimulus is off. }\\

\section{Funding Statement} 
We acknowledge support from PPG Grant P01 GM071862 from the National Institute of General Medical Sciences. This work was also supported by the D.R.Bullard-Welch Chair at Rice University, Grant C-0016 (to P.G.W). \\

\section{Acknowledgements}
Authors acknowledge the numerous fruitful discussions on $NF\kappa B$ biology with Prof Elizabeth Komives. 

\section{Author Contributions} 
ZW, DP and PGW designed the research. ZW, DP carried out the research. ZW, DAP and PGW wrote the manuscript. ZW and DAP have contributed equally to this work. \\

\subsection{}
\subsubsection{}

\bibliographystyle{apsrev}
\bibliography{PRL}
\end{document}